\documentclass[USenglish,pre,
10pt,twocolumn]{article}
\usepackage[latin1]{inputenc}
\usepackage{babel}
\usepackage{times}
\usepackage{vmargin}
\usepackage{graphicx}

\usepackage{amsmath}
\usepackage{amsthm}
\usepackage{amssymb}

\usepackage{enumerate}

\setpapersize{A4}
\setmarginsrb{2.5cm}{0.5cm}{2.5cm}{3.5cm}{1.5cm}{1cm}{1.5cm}{1.5cm}

\newcommand{\C}{\mathcal{C}}
\newcommand{\Cs}{\mathbf{C}} 
\newcommand{\SC}{\mathcal{S}}
\newcommand{\Ss}{\mathbf{S}}

\newcommand{\G}{\mathcal{G}}
\newcommand{\N}{\mathbb{N}}
\newcommand{\R}{\mathbb{R}} 
\newcommand{\ws}{{\wedge\star}}
\newcommand{\REL}{\mathcal{R}}

\newtheorem{definition}{Definition}

\title{Binding Social and Cultural Networks: A Model
\author{Camille Roth\footnote{CREA (Center for Research in Applied
    Epistemology), CNRS/Ecole Polytechnique, 1 rue Descartes, 75005 Paris, France.
    Corresponding author: \emph{roth@poly.polytechnique.fr}} , Paul
  Bourgine\footnotemark[\value{footnote}]}}
\begin{document}
\maketitle \abstract{\small Until now, most studies carried onto
  social or semantic networks have considered each of these networks
  independently.  Our goal here is to bring a formal frame for
  studying both networks empirically as well as to point out stylized
  facts that would explain their reciprocal influence and the
  emergence of clusters of agents, which may also be regarded as
  \emph{``cultural cliques''}. We show how to apply the Galois lattice
  theory to the modeling of the \emph{coevolution} of social and
  conceptual networks, and the characterization of cultural
  communities. Basing our approach on Barabasi-Albert's models, we
  however extend the usual preferential attachment probability in
  order to take into account the reciprocal influence of both
  networks, therefore introducing the notion of dual distance.  In
  addition to providing a theoretic frame we draw here a program of
  empirical tests which should give root to a more analytical model
  and the consequent simulation and validation.  In a broader view,
  adopting and actually implementing the paradigm of cultural
  epidemiology, once we have understood network formation and
  evolution, we could therefore proceed further with the study of
  knowledge diffusion and explain how the social network structure
  affects concept propagation and in return how concept propagation
  affects the social network.  }
  
\section*{Introduction}
Many studies have been carried on real networks, considering them as
complex systems and trying to explain their formation and dynamics
\cite{bara:stat,newm:rand}.  Whereas the models proposed have
initiated efficient proposals for explaining the general properties of
these networks (mostly about node degrees and in particular the
broadly shared \emph{scale-free} property \cite{bara:scal,vazq:know}),
yet they often lack robust explanations for ``advanced'' topological
features such as clustering \cite{newm:clus} -- a feature especially
observed in social networks which denotes the propensity of two agents
to be connected together if they have common acquaintances
\cite{stro:coll}.  Until recently though, no attempt had been made to
treat differently social networks in respect of other real networks.

A recent study by Newman \& Park \cite{newm:whys} however points out social
networks singularities regarding the correlations in degrees of
adjacent vertices as well as the clustering structure.  Trying to
model the way beliefs propagate among social networks of agents (in
our case, a community of scientists), that is, explain how the social
network structure affects concept propagation and in return how
concept propagation affects the social network, we will consider here
another approach stemming from a social psychology argument:
attraction for same-profile people (``homophily'') is indeed key in
the formation of social acquaintances \cite{mcph:homo}. 

Apart from properties relative to the social network such as node
degrees, we can assume that the dominant criterion for choosing a
scientific partner mostly depends on the cultural similarity of two
agents.  An economic model of knowledge creation developed in
\cite{cowa:join} already tries to take into consideration agents
profile (elements of a vector space) in order to explain the structure
of the economic network -- agents match two by two to produce new
knowledge according to their profile.

We introduce here a network dual to the social network, the network of
cultural representations, denoted as concepts.  Our goal is to
bring a formal frame for studying both networks empirically as well as
to point out stylized facts that would explain their reciprocal
influence and the emergence of clusters of agents, which may also be
regarded as \emph{``cultural cliques''}. In a broader view, adopting
and actually implementing the paradigm of cultural
epi\-de\-mio\-lo\-gy \cite{dawk:self,sper:cont}, we could therefore
proceed further with the study of knowledge diffusion.

\section{Networks}
  
We present hereafter the networks we work with -- social and
conceptual -- as well as the links within and between them.

\subsection{Social network}

\begin{definition}
  The social network $\SC$ is represented by the network of
  coauthorship, where nodes are authors and links are
  collaborations.
\end{definition}

Thus $\SC=(\Ss,\lambda_\Ss)$, where $\Ss$ denotes the set of authors and
$\lambda_\Ss$ denotes the set of undirected links.  As time evolves, new
articles are published, new nodes are possibly added to $\Ss$ and new
links are created between each pair of co-authors. We actually
consider the temporal series of networks $\SC(t)$ with $t\in \N$ 
(articles are published with a date, thus an integer), for
we want to observe the dynamics of the network. Yet we will usually
omit the reference to $t$ because $\SC$ always depends implicitly on
time.
\par
An important question of design here is the nature of links. Depending
on the model goals and precision, we may want to take into
account the fact that two nodes have co-authored more than one paper
(thus introducing \emph{link strength}), or that their collaborations
are more or less recent (thus introducing \emph{link age}). Indeed, an
empirical study of paper citation distribution \cite{redn:howp} shows
that the probability of citation decreases in respect of time, since papers are
gradually forgotten or obsolete; while another model examining
the world wide web network \cite{hube:evol} notices that the link
distribution must depend on the time that has elapsed since a web site
was created.

\paragraph{Weighted networks}
Relationships should consequently be different according to whether
agents have collaborated only once and a long time ago, or they have
recently co-authored many articles.  An easy and practical way for
dealing with these notions is to use a weighted network:
\begin{itemize}
\item in a \emph{non-weighted network}, we say that two nodes are
  linked as soon as there exists one coauthored article.  Links can
  only be active \emph{or} inactive.
\item in a \emph{weighted network}, links are provided with a weight
  $w \in \R^+$, possibly evolving in time. We can therefore easily
  represent multiple collaborations by increasing the weight of a
  link, or render the age of a relationship by decreasing this weight
  (for instance by applying an aging function). 
  
  This method enables us to model a non-weighted network by
  assigning weights of $1$ or $0$ respectively to active or inactive
  links.  This method also leaves room for creating \emph{ex post} a
  non-weighted network from a weighted network by setting a threshold,
  such that a link is active when its weight exceeds the threshold,
  otherwise inactive.
\end{itemize}

\subsection{Conceptual network}
\label{par:concnet}
The conceptual network is very similar to the social network - and as
we will see, dual:

\begin{definition}
  The conceptual network $\C$ is the network of joint
  appearances of concepts within articles, where nodes are concepts
  and links are co-occurrences.
\end{definition}
Identically to $\SC$, we have $\C=(\Cs,\lambda_\Cs)$.  When a new article
appears, new concepts are possibly added to the network, and new links
are added between co-appearing concepts. Here again, as in the case of
the social network, one may use a weighted network to render the
frequency or the age of co-occurrences.

However, the whole point is now to define precisely what a
\emph{concept} is. Is it a paradigm like \emph{``universal
  gravitation''}, a scientific field like \emph{``molecular
  biology''}, or a simple word like \emph{``interferon''}?  In
particular, what is a concept such that we can observe its appearance
in an article?

This notion needs be not too precise nor too wide. For instance,
authors provide their articles with keywords: apparently, considering
these keywords as concepts seems to constitute a relevant level
of categorization while being a convenient idea.  However, such
keywords have not proven to be very reliable indicators of the issues
articles are dealing with, for authors often omit important keywords
or specify poorly relevant ones.

\paragraph{Words as concepts} 
The idea would be to create new keywords from the words appearing in
articles, and at first we will say that \emph{each word is a concept}.
This definition does not prevent us from observing higher-level
concepts such as scientific fields or paradigms, since we can easily
refer to these concepts \emph{a posteriori} by considering sets of
strongly connected words. For example, we could interpret the set of
frequently co-occuring words \emph{\{``cell'', ``cancer'', ``DNA'',
  ``gene'', ``genetic'', ``genetics'', ``molecular''}\} as
\emph{molecular biology}. This understanding refers to the notion of
\emph{meme} introduced by Dawkins \cite{dawk:self}.

Moreover, we proceed only with words present in what we consider to be
the most relevant article data: the title and the abstract. We prefer
to set aside article content, since first and above all it is rarely
available; second, it could make appear too many very precise though
unrelevant words.  Therefore, we assume that all important concepts an
article tackles and bears on are explicitly used in its title or its
abstract. Of course, we also need define a list of words to be
ignored, or \emph{``stop words''}, including grammatical and
unsignificant words (\emph{``is", ``with", ``study",} etc.) as well as
non-discriminating words (e.g., \emph{``biology''} within a community
of biologists) for which a robust criterion will be proposed in
§\ref{par:apply}.

\subsection{Binding the two networks}
\label{par:binding}
As the social network is the network of joint appearances of authors,
so is the conceptual network with concepts, establishing an obvious
duality between the two networks. This duality is key if we want to
bind them and explain their reciprocal influence.

In the same way we did with the previous networks, we link scientists
to the words they use, i.e. we add a link whenever an author and a
word co-appear within an article.

Hence considering the two networks $\SC$ and $\C$, we deal with three
kinds of quite similar links: (i) between pairs of scientists, (ii)
between pairs of concepts, \hbox{and (iii)} between concepts and scientists;
thus setting up three kinds of binary relations:
\begin{enumerate}[(i)]
\item a set of symmetrical relations $\REL^\Ss_\alpha\:\subset \Ss\times
  \Ss$ from the social network to the social network, and such that
  given $\alpha\in\R$ and two scientists $s$ and $s'$, we have
  $\:s\;\REL^\Ss_\alpha\;s'\:$ \emph{iff} the link between $s$ and $s'$
  has a weight $w$ strictly greater than the threshold $\alpha$.
  
\item a set of symmetrical relations
  $\REL^\Cs_\alpha\:\subset \Cs\times \Cs$ from the conceptual network to
  the conceptual network, and such that given $\alpha\in\R$ and two
  concepts $c$ and $c'$, $\:c\;\REL^\Cs_\alpha\;c'\:$ \emph{iff} the
  link between $c$ and $c'$ has a weight $w>\alpha$.

\item a binary relation $\REL_\alpha\subset \Ss\times \Cs$ from the
  social network to the conceptual network, and such that given
  $\alpha\in\R$, an author $s$ and concept $c$,
  $\:s\;\REL_\alpha\;c\:$ \emph{iff} the link between $s$ and $c$ has
  a weight $w>\alpha$.
\end{enumerate}

Let us examine the special case $\alpha=0$. No\-ti\-cing that
$\alpha<\alpha'\Rightarrow\REL^{(.)}_{\alpha'}\subset\REL^{(.)}_\alpha$,
thus giving
${\forall\alpha>0},{\REL^{(.)}_\alpha\subset\REL^{(.)}_0}$, we infer
that the relations $\REL^{(.)}_0$ are maximal, i.e. two nodes are
related whenever there exists a link binding them, whatever its
weight. To ease the notation, we will identify $\REL^\Ss_0$ to $\REL^\Ss$,
$\REL^\Cs_0$ to $\REL^\Cs$, and $\REL_0$ to $\REL$.

\begin{figure}
\begin{center}
\includegraphics{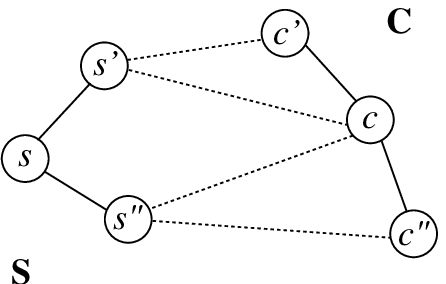}
\caption{Sample network with $\Ss=\{s,s',s''\}$, $\Cs=\{c,c',c''\}$, and
  relations $\REL^\Ss$, $\REL^\Cs$ (solid lines) and $\REL$ (dashed
  lines).}
\end{center}
\label{fig:reseau}
\end{figure}

\section{Lattices and epistemic closure}

The basic ingredients being defined, we need yet another formal tool
to formulate stylized facts about knowledge and people conveying it.
Galois lattices appear to be a suitable frame to describe these facts
as they offer a powerful structure for concept categorization. 
They are also being therefore widely used in conceptual knowledge systems
\cite{will:con2} and formal concept classification \cite{godi:meth}.
In the field of social networks, White \& Freeman have already explored an
application of this theory to social networks \cite{free:usin}, though
his model deals with agents and social events they attend. The goal of
this section is to present the Galois lattice theory and show how we can use it
here to describe efficiently the relationships between $\SC$ and $\C$.

\subsection{Sets and relations}
Let us first consider two finite sets $A$ and $B$ between which we
have a binary relation ${R \subseteq A\times B}$. We introduce the
operation ``$\wedge$'' such that for any element $x\in A$, $x^\wedge$
is the set of $B$ elements $R$-related to $x$. Extending this
definition to subsets $X\subseteq A$, we denote by $X^\wedge$ the set
of $B$ elements $R$-related to every element of $X$, namely:
\begin{subequations}\begin{align}
x^\wedge&=\{\,y\in B\;|\;x R y\,\}\\
X^\wedge&=\{\,y\in B\;|\; \forall x\in X, x R y\,\}
\end{align}\end{subequations}

Similarly, ``$\star$'' is the dual operation so that ${\forall y\in
  B}$, ${\forall Y\subseteq B}$,
\begin{subequations}\begin{align}
y^\star&=\{\,x\in A\;|\;x R y\,\}\\
Y^\star&=\{\,x\in A\;|\; \forall y\in Y, x R y\,\}
\end{align}\end{subequations}
By definition we set $(\emptyset)^\wedge=B$ and
$(\emptyset)^\star=A$. 

These operations enjoy the following properties:
\begin{subequations}\begin{align}
\label{eq:troisa}X\subseteq X' &\Rightarrow {X'}^\wedge\subseteq {X}^\wedge\\
Y\subseteq Y' &\Rightarrow {Y'}^\star\subseteq {Y}^\star\end{align}\end{subequations}
and
\begin{subequations}\begin{flalign}
X&\subseteq {X^\wedge}^\star\label{eq:quatrea}\\
Y&\subseteq {Y^\star}^\wedge\label{eq:quatreb}
\end{flalign}\end{subequations}
Also, we have\footnote{
  And accordingly,
  $
  X^\wedge=(\bigcup_{x\in X} \{x\})^\wedge=\bigcap_{x\in X}x^\wedge$.}:
\begin{subequations}\label{eq:unioninter}\begin{align}
(X\cup X')^\wedge = {X}^\wedge\cap{X'}^\wedge\\
(Y\cup Y')^\star = {Y}^\star\cap{Y'}^\star
\end{align}\end{subequations}

\paragraph{Closure operation}
More important, the following property holds true,\footnote{Indeed,
  (\ref{eq:troisa}) applied to (\ref{eq:quatrea}) leads to
  $(X^{\wedge\star})^\wedge \subseteq X^\wedge$, while
  (\ref{eq:quatreb}) applied to $X^\wedge$ gives $(X^\wedge) \subseteq
  (X^\wedge)^{\star\wedge}$.}
\begin{equation}
((X^\wedge)^\star)^\wedge=X^\wedge\:\text{ and }\:((Y^\star)^\wedge)^\star=Y^\star
\end{equation}
and therefore we are enabled to define the operation ``$\ws$'' as a
\emph{closure operation} \cite{birk:latt}, in that it is:
\begin{subequations}\begin{flalign}
    \text{extensive,}&&X&\subseteq X^\ws\\
    \text{idempotent}&&(X^\ws)^\ws&=X^\ws\label{eq:idem}\\
    \text{and increasing.}&&X\subseteq X' &\Rightarrow X^\ws\subseteq
    X'^\ws
\end{flalign}\end{subequations}
We say that $X$ is a \emph{closed} subset if $X^\ws=X$.

\subsection{Galois lattices}
We need now consider the set of couples of subsets of $A$ and $B$ and
build a new structure onto it.
\paragraph{Complete couples} 
Given two subsets $X\subseteq A$ and $Y\subseteq B$, a couple $(X,Y)$
is said to be \emph{complete} \emph{iff} $Y=X^\wedge$ and $X=Y^\star$.

Yet such a couple is actually a $(X,X^\wedge)$ where ${X^\ws=X}$.
Therefore, complete couples correspond obviously to couples of
subsets of $A$ and $B$ closed under $\ws$.
This will allow us to define a new kind of lattice from $A$, $B$ and
$R$. We first recall the definition of a \emph{lattice}:
\begin{definition}
  A set $(L, \sqsubseteq, \sqcup, \sqcap)$ is a \emph{lattice} if
  every finite subset $H\subseteq L$ has a \emph{least upper bound}
  in $L$ noted $\sqcup H$ and a \emph{greatest lower bound} in $L$ noted 
  $\sqcap H$ under the partial-ordering relation $\sqsubseteq$.
\end{definition}
In this respect the set of subsets of a set $X$ provided with the
usual inclusion, union and intersection, $(\mathcal{P}(X), \subseteq,
\cup, \cap)$, is a lattice. Any partially-ordered finite set is also a
lattice, and so is a \emph{Galois lattice} \cite{barb:ordr}:

\begin{definition}
  Given a relation $R$ between two finite sets $A$ and $B$, the
  \emph{Galois lattice} $\G_{A,B,R}$ is the set of every
  \emph{complete} couple $(X,Y)\subseteq A\times B$ under relation
  $R$. Thus,
  \begin{equation}\G_{A,B,R}=\{(X^\ws,X^\wedge)|X\subseteq A\}\end{equation}
\end{definition}

Indeed $\G_{A,B,R}$ is finite and is provided with the following
natural partial order $\sqsubseteq$:
\begin{equation}
(X,X^\wedge)\sqsubseteq(X',{X'}^\wedge) \Leftrightarrow X\subset X'
\end{equation}

\paragraph{Formal concept lattice} As Wille points out in \cite{will:con2},
this structure constitutes a solid formalization of the philosophical
apprehension of a concept characterized by its \emph{extension} (the
physical implementation or the group of things denoted by the concept)
and its \emph{intension} (the properties or the internal content of
the concept).

In a pair $g=(X,X^\wedge)$ considered as a formal concept, $X$ may be
seen as the extension of $g$ while $X^\wedge$ is its intension. For a
given $X\subseteq A$, $X^\wedge$ will represent the set of properties
shared by all objects of $X$, whereas for a given set of properties
$Y\subseteq B$, $Y^\star$ will be the set of objects of $A$ actually
fulfilling them.

Also, using the strict partial order $\sqsubset$, we can talk of
\emph{formal subconcept} by saying \emph{$g$ is a subconcept of $g'$
  iff $g\sqsubset g'$}. Hence $g$ can be seen as a specification of
$g'$, since the number of its properties increases ($X^\wedge\supset
X'^\wedge$, thus defining $g$ more precisely) while less objects
belongs to its extension ($X\subset X'$).  Conversely, $g'$ is a
\emph{``superconcept''} or a generalization of $g$; we have so a tool
of generalization and specification of formal concepts
\cite{will:conc}.\footnote{Of course, these notions are dually
  defined, i.e. it is possible to consider $Y$ as an extension and
  $Y^\star$ as an intension.}

\subsection{Applying lattices to $\SC$ and $\C$}
\label{par:apply}
It is possible now to give some semantics to these tools with respect
to our networks $\SC$ and $\C$.  For this purpose, we will consider the two
finite sets $\Ss$, $\Cs$, the relation $\REL$ and $\G_{\Ss,\Cs,\REL}$.

First, for an author $s\in \Ss$, $s^\wedge=\{\;c\;|\;s \REL c\;\}$
represents the set of the concepts he talked about or the fields he
dealt with. Proceeding identically with a concept $c\in \Cs$,
$c^\star=\{\;s\;|\;s \REL c\;\}$ represents the set of scientists who
used the concept $c$ in at least one of their papers. 

Then, for a group of authors $S\subseteq\Ss$, $S^\wedge$ represents
the words being used by every author $s\in S$, while for a set of
words $C\subseteq\Cs$, $C^\star$ is the set of agents using every concept
$c\in C$. Moreover, we can easily derive from (\ref{eq:unioninter})
the words used by a community $S\cup S'$ by taking the intersection
$S^\wedge\cap S'^\wedge$, or the authors corresponding to the merger
of any two sets of concepts $C\cup C'$ by taking $C^\star\cap
C'^\star$. 

\begin{figure}
\begin{center}
\includegraphics{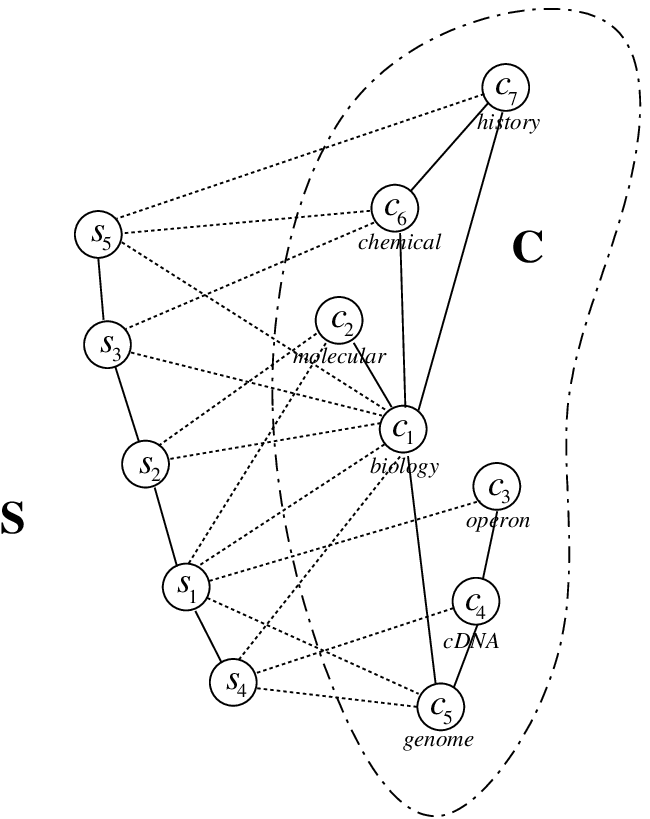}
\caption{\label{fig:reseau2}$\Ss=\{s_1,s_2,s_3,s_4, s_5\}$, $\Cs=\{c_1,c_2,c_3,c_4,c_5,c_6,c_7\}=
  \{$biology, molecular, operon, cDNA, genome, chemical,
    history$\}$ -- solid lines: $\REL^\Ss$ and $\REL^\Cs$, dashed
  lines: $\REL$. }
\end{center}
\end{figure}

An example is shown on figure \ref{fig:reseau2}.  For instance,
${s_4}^\wedge=\{c_1,c_4,c_5\}$ and $\{c_1,c_6\}^\star=\{s_3,s_5\}$.  If
we consider the matrix $R$ representing relation $\REL$ as follows,
\[R=\begin{pmatrix}
  1 & 1 & 1 & 0 & 1 & 0 & 0\\
  1 & 1 & 0 & 0 & 0 & 0 & 0\\
  1 & 0 & 0 & 0 & 0 & 1 & 0\\
  1 & 0 & 0 & 1 & 1 & 0 & 0\\
  1 & 0 & 0 & 0 & 0 & 1 & 1
\end{pmatrix}\]
where $R_{i,j}$ is non-zero when $s_i\;\REL\;c_j$, we can easily read
${s_i}^\wedge$ on rows and ${c_j}^\star$ on columns.

\paragraph{Epistemic closure and epistemic categories}
Seeing concepts as \emph{properties} of authors who use them (skills
in scientific fields as cognitive properties) and authors as
\emph{extensions} of concepts (implementation of concepts within
authors), one can make a very fertile usage of the lattice
$\G_{\Ss,\Cs,\REL}$ by setting up an epistemic taxonomy with the help
of formal concepts made of couples $(S,C)$ with $S\subseteq\Ss,
C\subseteq\Cs$. We may indeed consider such formal concepts as
\emph{schools of thought} constituted by the community of agents $S$
working and writing on the field $C$, a formal subconcept simply being
a trend inside a school. By community we understand henceforth
\emph{epistemic community}, that is to say neither a department nor a
group of research.
 
In addition, we recall that for such a complete couple from the Galois
lattice, $C=S^\wedge$, $S=C^\star$ and finally $S=S^\ws$.  What does
$S^\ws$ actually represent ? It is the set of scientists using
\emph{at least} the same words as $S$. But ``$\ws$'' being a closure
operation, $S^\ws$ closes the set $S$ by returning all the scientists
related to every concept shared among $S$ -- once and for all from
(\ref{eq:idem}) -- which makes us call it an \emph{epistemic closure
  operation}.\footnote{Note that given $S^\wedge=\{c_1,...,c_n, c\}$
  and $S'^\wedge=\{c_1,...,c_n, c'\}$, we have $S'\not\in S^\ws$, $S'$
  not being in the epistemic closure of $S$, which might look quite
  strange for a human eye who would have said their domains of
  interest similar.  
  
  Another property may help understand better what this closure
  actually tallies with: given $S^\wedge=\{c_1,...,c_n\}$ and
  $S'^\wedge=\{c'_1,...,c'_n\}$ such that $\forall
  (i,j)\in\{1,...,n\}^2,c_i\not = c'_j$, we have $(S\cup S')^\ws=\Ss$:
  the closure of two sets of scientists working on totally different
  issues is the whole community $\Ss$ -- ``there is no way to
  distinguish $S$ and $S'$ from each other with respect to the rest of
  the community''.}

Admittedly, for a single scientist $s$, $s^\ws$ will certainly be
equal to $s$, since there are strong chances that $\forall s'\in \Ss$,
$\exists w \in s^\wedge$ and $\not\in s'^\wedge$. Considering however
a subset $S\subseteq\Ss$, as its cardinal increases there are more and
more chances that the closure of $S$ reaches an actual community of
researchers.  We conjecture that there is a relevant level of closure
for which a set $S^\ws$, and identically $C^{\star\wedge}$, is
representative of a field or a trend.  This idea is to be compared to
Rosch's basic-level of categorization \cite{rosc:cogn}.

$\G_{\Ss,\Cs,\REL}$ contains all complete couples: this includes
naturally most singletons $(s^\ws,s^\wedge)$ as well as
$(\Ss,\Ss^\wedge)$, but also and especially all the intermediary pairs
of closed sets. For this purpose, there must then be a gap between
couples whose $S$ is of very small size and those with medium-sized
$S$, with very few complete couples inbetween; and likewise a gap
between couples with small- and medium-sized $C$ sets.  This medium
level shall constitute our basic-level of epistemic categorization,
whereas above it (``superordinate categories'') the field would be too
general, and too precise under it (``subordinate categories'').

If we define the \emph{epistemic family} of an agent as the set of
(possibly many) complete couple(s) which he is a member of, and also
whose $S$ set size is above a certain threshold, it could be very
useful to identify the basic-level epistemic categories to help fit
the threshold value.

\begin{figure}
\begin{center}
\includegraphics{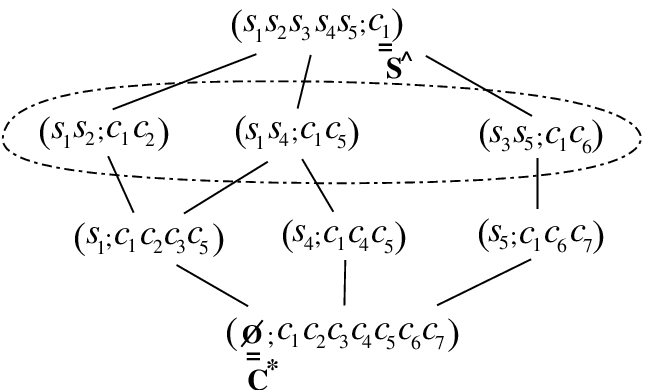}
\caption{\label{fig:reseau3}Representation of the whole Galois lattice of our
  example -- the hierarchy is drawn according to the partial order
  $\sqsubset$, i.e.  ``bottom''$\sqsubset$``top''. The cultural
  background $\Ss^\wedge$ is reduced to ``biology''.  On the
  medium-level, we find formal concepts $(s_1,s_2$ ; ``biology'' ,
  ``molecular''$)$, $(s_1,s_4$ ; ``biology'' , ``genome''$)$,
  $(s_3,s_5$ ; ``biology'' , ``chemical''$)$.}
\end{center}
\end{figure}

\paragraph{Cultural background} Interestingly, $\Ss^\wedge$ represents the
concepts the whole community shares -- the \emph{``background''} --
and are obviously too common to be discriminating.  This set could
actually constitute a very appropriate companion to the list of stop
words we mentioned in §\ref{par:concnet}. On the other hand,
$\Cs^\star$ does not enjoy in general any such property and is empty;
the contrary would mean that there would be at least one author having
used \emph{every} word in use among the whole community, which would be
quite dreadful in fact.

\section{Applications}

This section is devoted to pointing out the joint applications of the
two preceding sections to the observation, description and eventually
model of the dynamics of our networks.

\subsection{Network dynamics}

We will first try to account for the network evolution by extending
already existing models to make them include the improvements offered
by the theoretic frame exposed above. The growing-network model
proposed by Barabasi \& Albert \cite{bara:evol} will be our basis. It
is directed by two key phenomenas: (i) a \emph{constant rate of
  growth} (the number of nodes at any time $t$ is $\alpha t$),
justified by the fact that real networks ``grow by the continuous
addition of new nodes'' \cite{bara:stat}; and (ii) \emph{a
  preferential attachment} -- external (new nodes join the system) as
well as internal (links appearing between existing nodes) -- however
neglecting aging considerations.  This is borne out by the preference
one may  for example exhibit towards an already well-connected agent,
as being more recognized, famous, reliable or simply efficient.

If we assume yet that homophily is essential to the system dynamics,
the preferential attachment must be modified in order to take into
account similarity between agents or between concepts: nodes will
indeed join preferentially more connected but also more similar nodes.
Thus, the preferential attachment probability of a node to another
node within the same network, usually denoted by $\Pi(k_1,k_2)$ where $k_1,k_2$ are
the degrees of the nodes connecting to each other\footnote{This concerns 
internal attachment. External attachment is, of course, undefined as regards
the dual distance.}, should not be uniform with respect to the nodes proximity
within their dual network. That is to say for instance that $\Pi$
should depend for a scientist $s\in\Ss$ both \hbox{(i) on} the degree
of other scientists $s'\in\SC$ (using $\REL^\Ss$) and \hbox{(ii) on}
the \emph{``distance''} between $s^\wedge$ and $s'^\wedge$ -- or
\emph{dual distance} between $s$ and $s'$ (using $\REL^\Cs$ and
$\REL$).

\paragraph{Dual distance}The notion of dual distance $d$ needs nonetheless be
defined more precisely: in order to measure the similarity or
equivalently the difference between two nodes dual sets we will adopt
a formula inspired from the notion of Hamming distance on sets of
bits.
\begin{definition}Given $(s, s')\in \Ss^2$ and using 
  the symmetric difference of two sets we define the dual distance
  $d(s,s')\in[0;1]$ such that:\footnote{Written in a more explicit
    manner, given two sets $s^\wedge =
    \{c_1,...,c_n,c_{n+1},...,c_{n+p}\}$ and $s'^\wedge=
    \{c_1,...,c_n,c'_{n+1},...,c'_{n+q}\}$, $\displaystyle
    d(s,s')=\frac{p+q}{p+q+n}$; $n$ and $p$, $q$ representing
    respectively the number of elements $s^\wedge$ and $s'^\wedge$
    have in common and have in proper. We also verify that if $n=0$
    (disjoint sets), $d=1$; if $n\not =0$, $p=q=0$ (same sets), $d=0$;
    and if $s^\wedge\subset s'^\wedge$ (included sets),
    $d=\frac{q}{q+n}$. It is besides easy though cumbersome to show that $d(.,.)$ is 
    actually a distance.}
\[d(s,s')=\frac{|(s^\wedge \setminus s'^\wedge)
  \cup (s'^\wedge \setminus s^\wedge)|}{|s^\wedge\cup s'^\wedge|}\]
\end{definition}

We expect to find a different behavior with respect to this parameter
$d$: indeed if cliques do exist, preferential attachment should also
depend on a parameter precisely related to the
\emph{``cliquishness''}.  Newman in \cite{newm:clus} considers the
number of common acquaintances as an explanatory argument for clique
formation.  Instead, we will assume that collaborations do essentially
occur on account of homophily, while this assumption does not
contradict Newman's argument: two agents are all the more likely to
have the same profile that they share many
acquaintances.\footnote{However another yet better model would also
  take into account this property and would express it through
  enhancing our definition of $d$.}

The probability $\Pi(k_1,k_2)$ encloses this information without
enabling us to discriminate the effect of the dual proximity. We next
consider $\Pi(k_1,k_2,d)$ which takes into account this second
variable $d$. The main direction to be explored would be to build
$\Pi(k,d)$ on the dual distance defined above. Assuming the
independence of $(k_1,k_2)$ and $d$,
$\int_0^1\Pi(k_1,k_2,\delta)\rho(\delta)d\delta=\Pi(k_1,k_2)$ holds
true, where $\rho$ is the density of $d$.\footnote{We could also draw
  out the distribution $\Pi(k_1,k_2,d)$ depending on whether agents
  are member of the same epistemic family (introduced in
  §\ref{par:apply}) or not.  The variable $d$ would thus belong to
  $\{0;1\}$ and we would actually deal with only two distributions.
  This option is less robust than the previous one for it relies on a
  quite ``fuzzy'' parameter (epistemic family membership), whereas it
  could offer a more schematic and stylized interpretation.}

In any case, a first step will be to determine empirically the shape
of $\Pi$ so that we can infer fertile intuitions for designing an
analytical value for $\Pi$. Also, although not detailed here, the
reasoning holds the same for the preferential attachment in $\Cs$.

\subsection{\emph{Cliquishness} and coalitions}
As regards the cliquishness in particular, another point of interest
is to see whether network cliques correspond to closed sets, i.e.
whether a $\SC$-clique is also a $\ws$-clique, and whether a
$\C$-clique is also a $\star\wedge$-clique. In other words, we want to
know whether schools of thought and scientific fields are also
socially and conceptually strongly linked or not. Though we could
expect this to be true in real world networks, it is certainly not a
fortuitous property for it relies on two different kinds of tools
(epistemic closure vs.  single network connectivity). We might also
want to adopt here an extended definition of a clique (as a fully
connected triplet of nodes), and for instance use
\emph{$k$-connectivity}\footnote{We acknowledge fruitful remarks from
  Douglas R. White on this point.} \cite{powe:grow} (the smallest
number of nodes one needs to withdraw from a connected (sub)graph to
get a disconnected one).

\subsection{Sensibility to parameter $\alpha$}
At the beginning of our paper (§\ref{par:binding}) we left room for a
parameter $\alpha$ in relations $\REL_\alpha^{(.)}$, which
had been until now implicitly set to $0$. Its main function is actually to
 prevent unsignificant
links (too old or too rare) to be taken into account: indeed, under a
certain threshold of strength or significance, a link would be
excluded from $\REL_\alpha^{(.)}$. 

In the extreme case, for $\alpha$ big enough there is no connection at
all. As for the appearance of a giant component in random networks
\cite{newm:rand}, there may be a transition value $\alpha_c$ above
which almost no connectivity exists and under which the network is
significantly connected. This hypothesis ought also to be checked.

\section*{Conclusion}
Until now, most studies carried onto social networks or conceptual
(semantic) networks have considered each of these networks
independently. We proposed here a frame for binding them and pointing
out their very duality as well as expressing stylized facts about
them. The Galois lattice theory has proved useful in helping introduce
key notions such as epistemic closure and basic-level of
categorization of a scientific field, and in general for
characterizing scientific communities.

Next we showed how to apply this structure to the model of the
\emph{coevolution} of the social and cultural networks. We have mostly
based our approach on Barabasi-Albert's models, stressing out constant
growth and preferential attachment. However instead of considering
social and conceptual networks separately, we modified the usual
preferential attachment probability in order to take into account the
reciprocal influence of both networks. We therefore introduced the
notion of dual distance.

Finally, more than providing a theoretic frame, we have in fact also
drawn a program of empirical tests which should give root to a more
analytical model and the consequent simulation and validation.
This step constitutes the first milestone of a broader attempt to
implement the paradigm of cultural epi\-de\-mio\-lo\-gy: we could thus
describe and explain propagation of concepts through the social
network \emph{as well as} propagation of scientists through the
conceptual network.  \vspace{1cm} \hrule

 
\end{document}